	\documentclass[twocolumn,twoside]{IEEEtran}	

	\usepackage{changebar}
	\usepackage[T1]{fontenc}        
	\usepackage{xspace}
	\usepackage{theorem}
	\usepackage{bbold}
	\usepackage{citesort}
	\usepackage{amssymb}
	\usepackage{times} 
	\usepackage[final]{graphicx}
	\graphicspath{{./epsfig/}}	

	\def\sss{\scriptscriptstyle}
	\def\dispty{\displaystyle}
	
	\def\ML{^{\sss \mathrm{ML}}}
	
	\def\MAP{^{\sss \mathrm{MAP}}}
	\def\wrt{w.r.t.\xspace}
	\def\pdf{p.d.f.\XS}
	\def\prior{\textit{prior}\XS}
	\def\Prior{\textit{Prior}\XS}
	\def\Prob#1{\Pr\cro{#1}}
	
\RequirePackage{amsmath}
\RequirePackage{xspace}
\RequirePackage{bbm}

\def\XS{\xspace}
\DeclareMathAlphabet{\mathb}{OML}{cmm}{b}{it}
\def\sbm#1{\ensuremath{\mathb{#1}}}                
\def\sbmm#1{\ensuremath{\boldsymbol{#1}}}          
\def\sdm#1{\ensuremath{\mathrm{#1}}}               
\def\sbv#1{\ensuremath{\mathbf{#1}}}               
\def\scu#1{\ensuremath{\mathcal{#1\XS}}}           
\def\sbl#1{\ensuremath{\mathbbm{#1}}}              

  \def\ab{{\sbm{a}}\XS}
  \def\bb{{\sbm{b}}\XS}

  \def\kb{{\sbm{k}}\XS}

  \def\rb{{\sbm{r}}\XS}

  \def\ub{{\sbm{u}}\XS}

  \def\yb{{\sbm{y}}\XS}
  \def\zb{{\sbm{z}}\XS}

\def\Bc{{\scu{B}}\XS}   
\def\Cc{{\scu{C}}\XS}   
\def\Dc{{\scu{D}}\XS}   
   
\def\Fc{{\scu{F}}\XS}

\def\Lc{{\scu{L}}\XS}   
   
\def\Nc{{\scu{N}}\XS}   
   
\def\Pc{{\scu{P}}\XS}

\def\Yc{{\scu{Y}}\XS}   
\def\Zc{{\scu{Z}}\XS}

  \def\aD{{\sdm{a}}\XS}
  \def\bD{{\sdm{b}}\XS}

\def\HD{{\sdm{H}}\XS}

\def\MD{{\sdm{M}}\XS}  \def\mD{{\sdm{m}}\XS}

\def\TD{{\sdm{T}}\XS}  \def\tD{{\sdm{t}}\XS}

\def\Cbb{{\sbl{C}}\XS}

\def\Kbb{{\sbl{K}}\XS}

\def\Nbb{{\sbl{N}}\XS}  
\def\Obb{{\sbl{O}}\XS}  
\def\Pbb{{\sbl{P}}\XS}  \def\pbb{{\sbl{p}}\XS}
  
\def\Rbb{{\sbl{R}}\XS}

\def\Zbb{{\sbl{Z}}\XS}

\def\nub         {{\sbmm{\nu}}\XS}

\def\unb     {{\sbv{1}}\XS}        \def\unbb     {{\sbl{1}}\XS}

\def\eC{\Cbb}

\def\eN{\Nbb}

\def\eR{\Rbb}
\def\eZ{\Zbb}

\def\XS{\xspace}

\RequirePackage{ifthen}
\RequirePackage{calc}

\newcommand{\taille}[1][\scad]{%
\ifthenelse{#1 = -5}{}{}%
\ifthenelse{#1 = -4}{\tiny}{}%
\ifthenelse{#1 = -3}{\scriptsize}{}%
\ifthenelse{#1 = -2}{\footnotesize}{}%
\ifthenelse{#1 = -1}{\small}{}%
\ifthenelse{#1 = 0}{\normalsize}{}%
\ifthenelse{#1 = 1}{\large}{}%
\ifthenelse{#1 = 2}{\Large}{}%
\ifthenelse{#1 = 3}{\LARGE}{}%
\ifthenelse{#1 = 4}{\huge}{}%
\ifthenelse{#1 = 5}{\Huge}{}}

\newboolean{@serif}
\setboolean{@serif}{true}
\def\scad{-5} 

\newcounter{taille}
\newcommand{\sca}[2][\scad]{\setcounter{taille}{#1}%
  \ifthenelse{\boolean{@serif}}
  {{\taille[\thetaille]\textsc{#2}}}
  {\setcounter{taille}{\value{taille}-1}{\uppercase{\taille[\thetaille]#2}}}}

\def\rem#1{}                    

\def\apost{\textit{a posteriori}\XS}

\def\post{\textit{posterior}\XS}
\def\Post{\textit{Posterior}\XS}

\def\aprio{\textit{a priori}\XS}

\def\eg{\textit{e.g.,}\XS}

\def\etc{\textit{etc\dots}\XS}

\def\ie{\textit{i.e.,}\XS}

\def\wrt{w.r.t.\XS}

\def\ieme{$^{\text{\small e}}$\XS}

\newcommand{\cnd}[1][\scad]{\sca[#1]{cnd}\XS}

\newcommand{\dess}[1][\scad]{\sca[#1]{dess}\XS}
\newcommand{\dea}[1][\scad]{\sca[#1]{dea}\XS}

\newcommand{\eea}[1][\scad]{\sca[#1]{eea}\XS}

\newcommand{\td}[1][\scad]{\sca[#1]{td}\XS}
\def\TD{travaux dirigés\XS}
\newcommand{\tp}[1][\scad]{\sca[#1]{tp}\XS}
\def\TP{travaux pratiques\XS}

\newcommand{\gbm}[1][\scad]{\sca[#1]{gbm}\XS}

\newcommand{\irm}[1][\scad]{\sca[#1]{irm}\XS}

\newcommand{\pc}[1][\scad]{\sca[#1]{pc}\XS}

\newcommand{\beaunom}[4][0]{{\taille[#1]#2 \uppercase{#3}}%
{\sca[#1]{#4}}\XS}

\newcommand{\JFG}[1][\scad]{\beaunom[#1]{Jean--Fran\c{c}ois}{G}{iovannelli}}

\def\CdO{Centre d'Orsay\XS}

\newcommand{\cnrs}[1][\scad]{\sca[#1]{cnrs}\XS}

\newcommand{\inserm}[1][\scad]{\sca[#1]{inserm}\XS}

\def\LSS{Laboratoire des Signaux et Syst\`emes\XS}

\def\UPS{Universit\'e de Paris-Sud\XS}
\newcommand{\ups}[1][\scad]{\sca[#1]{ups}\XS}

\newcommand{\ESE}[1][\scad]{\sca[#1]{sup\'elec}\XS}

\newcommand{\adresse}[1][\scad]{\ESE[#1], Plateau de Moulon, 91192 Gif--sur--Yvette Cedex, France\XS}
\newcommand{\adresscea}[1][\scad]{4 Place du Général Leclerc, 91406 Orsay Cedex, France\XS}
\newcommand{\tutelle}[1][\scad]{\cnrs[#1]~--~\ESE[#1]~--~\ups[#1]}

\def\ALSSun{\LSS (\tutelle) \adresse}
\def\helvnormal#1{{\fontfamily{phv}\fontsize{11}{14pt}\selectfont #1}}
\def\helvfoot  #1{{\fontfamily{phv}\fontsize{9}{14pt} \selectfont #1}}
\def\helvscript#1{{\fontfamily{phv}\fontsize{8}{14pt} \selectfont #1}}
\def\pth#1{\left(#1\right)}                
              
\def\cro#1{\left[#1\right]}

\def\bigcro#1{\bigl[#1\bigr]}

\def\Exp#1{\exp\cro{#1}}                

\def\Det#1{\det\bigcro{#1}}     

\def\Pr{\mathop{\textrm{Pr}}}

\def\IF{\text{if\:}}             
             
\def\AND{\text{and\:}}

\def\FOR{\text{for\:}}

\def\WITH{\text{with\:}}

\newsavebox{\fminibox}
\newlength{\fminilength}
\newenvironment{fminipage}[1][\linewidth]
  {\setlength{\fminilength}{#1}
   \begin{lrbox}{\fminibox}\begin{minipage}{\fminilength}}
  {\end{minipage}\end{lrbox}\noindent\fbox{\usebox{\fminibox}}}

 \def\T{^\tD} \def\+{^\dagger}
\def\I{\,|\,}           

\def\nequiv{\not\kern-.05em\equiv}
\def\egal{\kern-.5em=\kern-.5em}        
\def\propt{\kern-.2em\propto\kern-.2em} 
\def\wh#1{\widehat{#1}}                 

\def\argmax{\mathop{\mathrm{arg\,max}}} 
\def\argmin{\mathop{\mathrm{arg\,min}}} 

\def\froc#1#2{{#1/#2}}                  

\def\intdouble{\int\kern-0.3em\int}
\def\inttriple{\int\kern-0.3em\int\kern-0.3em\int}

\def\rond#1{\overset{\kern-0.33em~_\circ}{#1}}
\def\rondit[#1]#2{\overset{\kern#1~_\circ}{#2}}

\def\babs{\begin{abstract}}             \def\eabs{\end{abstract}}
\def\barr{\begin{array}}                \def\earr{\end{array}}
\def\bcc{\begin{center}}                \def\ecc{\end{center}}
\def\cl#1{\centerline{#1}}
\def\bdes{\begin{description}}          \def\edes{\end{description}}
\def\bdoc{
\begin{document}}             \def\edoc{\end{document}}
\def\ben{\begin{enumerate}}             \def\een{\end{enumerate}}
\def\beqn{\begin{eqnarray}}             \def\eeqn{\end{eqnarray}}
\def\beqnl#1{\beqn\label{#1}}           \def\eeqnl#1{\label{#1}\eeqn}
\def\beqnx{\begin{eqnarray*}}           \def\eeqnx{\end{eqnarray*}}
\def\bseqn{\begin{subeqnarray}}         \def\eseqn{\end{subeqnarray}}
\def\beq#1\eeq{\begin{equation}#1\end{equation}}
\def\bal#1\eal{\begin{align}#1\end{align}}
\def\balx#1\ealx{\begin{align*}#1\end{align*}}
\def\beqx{$$}                           \def\eeqx{$$}
\def\bfig{\protect\begin{figure}}       \def\efig{\protect\end{figure}}
\def\bfigx{\protect\begin{figure*}}     \def\efigx{\protect\end{figure*}}
\def\bfigt{\protect\begin{figurette}}   \def\efigt{\protect\end{figurette}}
\def\bfl{\begin{flushleft}}             \def\efl{\end{flushleft}}
\def\bfr{\begin{flushright}}            \def\efr{\end{flushright}}
\def\bit{\begin{itemize}}               \def\eit{\end{itemize}}
\def\bmi{\begin{minipage}}              \def\emi{\end{minipage}}
\def\bfmi{\begin{fminipage}}            \def\efmi{\end{fminipage}}
\def\bpic{\begin{picture}}              \def\epic{\end{picture}}
\def\bqu{\begin{quote}}                 \def\equ{\end{quote}}
\def\bqun{\begin{quotation}}            \def\equn{\end{quotation}}
\def\bsl{\begin{slide}}                 \def\esl{\end{slide}}
\def\btabb{\begin{tabbing}}             \def\etabb{\end{tabbing}}
\def\btabl{\begin{table}}               \def\etabl{\end{table}}
\def\btablx{\begin{table*}}             \def\etablx{\end{table*}}
\def\btab{\begin{tabular}} 
\def\btabu{\begin{tabular}}             \def\etabu{\end{tabular}}
\def\btabx{\begin{tabular*}}            \def\etabx{\end{tabular*}}
\def\bbib{\begin{thebibliography}{999}} \def\ebib{\end{thebibliography}}
\def\bver{\begin{verbatim}}             \def\ever{\end{verbatim}}
\def\bca{\begin{cases}}                          \def\eca{\end{cases}}
%
%

\typeout{\space}
\typeout{\space\space\space\space Fichier 'defgio.tex' -- JFG}
\typeout{\space}

\def\dspsty{\displaystyle}
\def\sss{\scriptscriptstyle}


	\def\atis{\textsc{atis}\XS}
	\def\dess{\textsc{dess}\XS}
	\def\dea{\textsc{dea}\XS}
	\def\eea{\textsc{eea}\XS}
	\def\Fiupso{\textsc{Fiupso}\XS}
	\def\FiupsoI{\textsc{Fiupso-i}\XS}
	\def\FiupsoII{\textsc{Fiupso-ii}\XS}
	\def\FiupsoIII{\textsc{Fiupso-iii}\XS}
	\def\Maitrise{Ma\^\i trise\XS}

	\def\td{\textsc{td}\XS}
	\def\TD{travaux dirigés\XS}
	\def\GTD{Travaux dirigés\XS}

	\def\tp{\textsc{tp}\XS}
	\def\TP{travaux pratiques\XS}
	\def\GTP{Travaux pratiques\XS}

	\def\gbm{\textsc{gbm}\XS}
	\def\elq{électronique\XS}
	\def\Elq{\'Electronique\XS}
	\def\ssl{\textsc{ssl}\XS}
	\def\SSL{signaux et systèmes linéaires\XS}
	\def\GSSL{Signaux et systèmes linéaires\XS}

	\def\Unixbf{U{\footnotesize NIX}\XS}
	\def\Shellbf{S{\footnotesize HELL}\XS}
	\def\Unix{\textsc{Unix}\XS}
	\def\Linux{\textsc{Linux}\XS}
	\def\Shell{\textsc{Shell}\XS}
	\def\Korn{\textsc{Korn}\XS}
	\def\mtlb{\textsl{matlab}\XS}
	\def\Mtlb{\textsl{Matlab}\XS}
	\def\pc{\textsc{pc}\XS}

	\def\msa{\medskipamount}
	\def\Prog#1{
	\texttt{ 
	\indent\indent\indent
	\begin{minipage}{15cm}
	\begin{tabbing} 
	~~~~~ \= ~~~~~ \= ~~~~~ \= ~~~~~ \= ~~~~~\=
	\\
	#1
	\end{tabbing}
	\end{minipage}
	}	
	}	

	\def\UPSCdO{
	\begin{small}
	\begin{tabular}{c} 
	\UPS	\\ 
	\CdO	\\ 
	-----	\\ \\ \\
	\end{tabular} 
	\end{small}
	}

	\def\Filiere#1{
	\begin{small}
	\begin{tabular}{c} 
	#1						\\ 
	Année 2001-2002	\\ 
	-----					\\ \\ \\
	\end{tabular} 
	\end{small}
	}

	\def\LogoUPS{\includegraphics[height=1.25cm,bb=0 421 596 842]{LogoCdO}}

\def\HeadUPS#1{
	\noindent
	\bcc
	\begin{tabular*}{\textwidth}{l@{\extracolsep\fill}r} 
	\UPSCdO	& \Filiere{#1}	\\ 
	\end{tabular*} 

	\vspace*{-1.75cm}
	\LogoUPS
	\ecc
}

	\def\Signature{
	\vfill
	\begin{small}
	\begin{flushright}
	\JFG \\
	Le \today
	\end{flushright}
	\end{small}
	}


	\def\Fcs{$^{\rm \,F}$\XS}
	\def\ieme{$^{\rm e}$\XS}
	\def\rib{\textsc{rib}\XS}
	\def\bp{\textsc{bp}\XS}

	\def\TSVP{
	\vfill
	\begin{flushright}
	\dots/\dots
	\end{flushright}
	\pagebreak
	}

	\def\Trait{\cl{\rule{2cm}{0.02cm}}}

	\def\BibTeX{{\rm B\kern-.05em{\sc i\kern-.025em b}\kern-.08em
				T\kern-.1667em\lower.7ex\hbox{E}\kern-.125emX}}

	\def\frem{\hfill$\triangle$}
	\def\freq{fréquence\XS}
	\def\irm{\textsc{irm}\XS}
	\def\KG{Kitagawa \& Gersch\XS}
	\def\perio{périodogramme\XS}
	\def\WV{Vigner--Ville\XS}
	\def\cpq{c'est pourquoi\XS}
	\def\Cpq{C'est pourquoi\XS}

	\def\dsp{\textsc{dsp}\XS}
	\def\tf{\textsc{tf}\XS}
	\def\fft{\textsc{fft}\XS}
	\def\ifft{\textsc{ifft}\XS}
	\def\fftdd{\textsc{fft-2d}\XS}

	\def\dimtext#1#2#3#4#5#6#7#8#9{
	\setlength{\textwidth}{#1}
	\setlength{\textheight}{#2}
	\setlength{\oddsidemargin}{#3}
	\setlength{\evensidemargin}{#4}
	\setlength{\topmargin}{#5}
	\setlength{\headheight}{#6}
	\setlength{\headsep}{#7}
	\setlength{\voffset}{#8}
	\setlength{\hoffset}{#9}
	}


	\def\pmu{^{-1}}

	\def\dpdx#1#2{{{\partial {#1}\over \partial {#2}}}}

	\def\est#1{\hat{#1}} 

	\def\Prob#1{\Pr{\left[#1\right]}}

	\def\fxsp#1#2{ f \left(  #1 \vert #2 \right) }

	\def\TFde#1{\Fc\cro{#1}\XS}
	\def\TLde#1{\Lc\cro{#1}\XS}
	\def\TZde#1{\Zc\cro{#1}\XS}

	\def\tf{\textsc{tf}\XS}
	\def\sf{\textsc{sf}\XS}
	\def\tl{\textsc{tl}\XS}
	\def\tz{\textsc{tz}\XS}
	\def\tpsc{\textsc{tc}\XS}
	\def\tpsd{\textsc{td}\XS}

	\def\TF{transformée de Fourier\XS}
	\def\SF{série de Fourier\XS}
	\def\TL{transformée de Laplace\XS}
	\def\TZ{transformée en $z$\XS}
	\def\TpsC{temps continu\XS}
	\def\TpsD{temps discret\XS}
	\def\RdC{région de convergence\XS}

	\def\GTF{Transformée de Fourier\XS}
	\def\GSF{Série de Fourier\XS}
	\def\GTL{Transformée de Laplace\XS}
	\def\GTZ{Transformée en $z$\XS}
	\def\GTpsC{Temps continu\XS}
	\def\GTpsD{Temps discret\XS}
	\def\GRdC{Région de convergence\XS}

	\def\VA{variable aléatoire\XS}
	\def\VAs{variables aléatoires\XS}
	\def\va{\textsc{va}\XS}

	\def\DSP{densité spectrale de puissance\XS}
	\def\dsp{\textsc{dsp}\XS}

	\def\TDS{traitement du signal\XS}
	\def\GTDS{Traitement du signal\XS}
	\def\tds{\textsc{tds}\XS}

	\def\CND{contrôle non-destructif\XS}
	\def\GCND{Contrôle non-destructif\XS}
	\def\cnd{\textsc{cnd}\XS}

\newcommand{\Entete}[5][Nom.Pr\'enom]{
\helvscript{\renewcommand{\arraystretch}{.8}
\begin{tabular}{c} Supélec
\\--------\end{tabular} \hfill
\begin{tabular}{c}Centre national de la recherche scientifique          \\
--------\end{tabular} \hfill
\begin{tabular}{c}\UPS  \\--------\end{tabular}} \\[-.4cm]%
\begin{center}\helvnormal{\textbf{LABORATOIRE DES SIGNAUX \& SYST\`EMES}}\\%
\helvfoot{Unit\'e mixte de recherche n$^{\rm o}$ 8506\\%
Sup\'elec, plateau de Moulon, 
3 rue Joliot-Curie, 91192 \sca[-3]{Gif--sur--Yvette} Cedex (France)\\%
T\'el\'ephone\,: 01 69 85 17 12 --- 
T\'el\'ecopie\,: 01 69 85 17 65 --- 
Courriel\,: {#1}@lss.supelec.fr}%
\\[0.5cm]\end{center}%
} 

\newcommand{\Inhead}[5][Name.Surname]{
\helvscript{\renewcommand{\arraystretch}{.8}
\begin{tabular}{c} Supélec
\\--------\end{tabular} \hfill
\begin{tabular}{c}Centre national de la recherche scientifique  \\
--------\end{tabular} \hfill
\begin{tabular}{c}\UPS  \\--------\end{tabular}} \\[-.4cm]%
\begin{center}\helvnormal{\textbf{LABORATOIRE DES SIGNAUX \& SYST\`EMES}}\\%
\helvfoot{Unit\'e mixte de recherche n$^{\rm o}$ 8506\\%
Sup\'elec, plateau de Moulon, 
3 rue Joliot-Curie, 91192 \sca[-3]{Gif--sur--Yvette} Cedex (France)\\%
Telephone: 01 69 85 17 12 --- 
Fax: 01 69 85 17 65 --- 
E-mail: {#1}@lss.supelec.fr}%
\\[0.5cm]\end{center}%
}

	\addtolength{\headheight}{0cm}
	\addtolength{\headsep}{0cm}
	\addtolength{\topmargin}{-0.50cm}

	\newtheorem{remark}{Remark}
	\newtheorem{proposition}{Proposition}
	\newtheorem{definition}{Definition}

	\def\dispsty{\displaystyle}

	\setlength{\fboxsep}{0.3cm}

	\def\LargeBox{8.2cm}
	
\bdoc
\setboolean{@serif}{false}

	\title{Unsupervised Frequency Tracking beyond the Nyquist Frequency using Markov Chains}

\author{Jean-Fran\c{c}ois Giovannelli, J\'er\^ome Idier, Redha Boubertakh and Alain Herment
\thanks{Jean-Fran\c{c}ois Giovannelli (giova@lss.supelec.fr) and J\'er\^ome Idier are
with the \ALSSun. Redha Boubertakh and Alain Herment are with the Unit\'e \inserm 494,
Imagerie M\'edicale Quantitative, H\^opital de la Piti\'e Salp\'etri\`ere, 91 Boulevard
de l'H\^opital, 75013 Paris.}
}

\markboth
{IEEE Transactions on Signal Processing}{Giovannelli \textit{\lowercase{et al.}}: Frequency Tracking beyond the Nyquist Frequency using Markov Chains}

\maketitle

\begin{abstract} 
This paper deals with the estimation of a sequence of frequencies from a corresponding
sequence of signals. This problem arises in fields such as Doppler imaging where its 
specificity is twofold. First, only short noisy data records are available (typically four
sample long) and experimental constraints may cause spectral aliasing so that measurements
provide unreliable, ambiguous information. Second, the frequency sequence is smooth. Here, this
information is accounted for by a Markov model and application of the Bayes rule yields the
\apost density. The maximum \apost is computed by a combination of Viterbi and descent
procedures. One of the major features of the method is that it is entirely unsupervised.
Adjusting the hyperparameters that balance data-based and \prior-based information is done
automatically by ML using an EM-based gradient algorithm. We compared the proposed
estimate to a reference one and found that it performed better: variance was greatly
reduced and tracking was correct, even beyond the Nyquist frequency.
\end{abstract}

\begin{keywords} 
Frequency tracking, aliasing inversion, regularization, Bayesian statistic, maximum \apost,
Viterbi algorithm, hyperparameter estimation, maximum likelihood, EM algorithm,
Forward-Backward procedure, ultrasonic Doppler velocimetry, meteorological Doppler radar.
\end{keywords}


\section{Introduction}

\PARstart{F}{requency tracking} (or mean frequency tracking) is currently of
interest~\cite{Boashash92a,Boashash92b,Barret93,Tichavsky97,Kootsookos98,So00}, especially 
in fields such as the ultrasonic characterization of biological tissues, synthetic aperture
radar, and speech processing. Our main interest is its use in Doppler imaging
(radars~\cite{Dias00}, ultrasound blood flow
mapping~\cite{Herment93,Giovannelli94a,Hann99}). There are two main features in this area. 

\ben
\item One is that only short noisy data records are available (typically four sample long) and they are in a vectorial form. Moreover, the constraints on the sampling frequency
may cause spectral aliasing, so that measurements provide small amounts of ambiguous
information. 
\item The second is that there is information on the smoothness of the sought frequency
sequence. This \aprio information is the foundation of the proposed construction. It allows
robust tracking, even beyond the Nyquist limit.
\een

The most popular methods used for spectral characterization rely on periodogram and
empirical correlations. The mean frequency is usually estimated by computing the mean
frequency of the periodogram~\cite{Herment93} over the standardized frequency range
$\nu\in(-0.5,+0.5]$. Another popular estimate is proportional to the phase of the first
empirical correlation lag~\cite{Kasai85,Woodman85}. It is also provided by a first-order
autoregression in a least squares framework~\cite{Loupas90}. But better accuracy is
obtained by using all the available estimated correlation lags in a Taylor series expansion
of the correlation function~\cite{Angelsen83,Woodman85}. The resulting estimate is also the
mean frequency of the periodogram. However, the estimated parameters vary greatly,
particularly when short data records are used.  Moreover, the estimated frequency
approaches zero when the true frequency becomes near the Nyquist frequency $\nu \simeq \pm
0.5$, (due to the periodogram 1-periodicity)~\cite{Herment93}. To reduce this bias,
\cite{Li85} uses the maximum of the periodogram instead of its mean (and yields a maximum
likelihood estimate, see Section~\ref{ProposedLikelihood} and~\cite[p.410]{Kay88}), 
and~\cite{Herment93} iteratively shifts the frequency of the data. This results in greater
variance so that no frequency tracking remains possible beyond $\nu=\pm 0.5$. 

Thus, all the current methods have two drawbacks. First, 
the tracking problem is tackled by a (necessary sub-optimal) two-step procedure: 
\ben

\item estimate frequencies in the aliased band $(-0.5,+0.5]$;

\item detect and inverse aliasing.

\een
Second, 
they are clearly based upon empirical second-order statistics that perform poorly with short data records independently processed. 
Unfortunately, the inverse aliasing in step~2 often fails due to the great variations in
the estimated aliased frequencies of step~1. This is usually compensated for by
post-smoothing the aliased frequency sequence. This provides spatial continuity but affects
the aliased frequency discontinuities, so limiting the capacity to detect aliasing.  The
proposed method copes with the great variation and aliasing in a single step; it models the
whole data set (by noisy cisoids) and the smoothness of the frequency sequence (by a Markov
random walk) in the regularization/Bayesian framework. It then becomes possible to smooth
frequency sequence and invert aliasing at the same time, so avoiding the pitfalls of
chaining these operations. 

We have found few papers~\cite{Streit90,Barret93,Chornoboy93} that adopt such a framework,
and this study provides four additional features. 

\ben

\item First, it deals with vectorial data records as they occur in Doppler imaging (see
Section~\ref{Statement}).

\item Second, it enables tracking beyond the Nyquist frequency, whereas others have not
investigated this problem. 

\item Third, exact frequency likelihood functions are computed
whereas~\cite{Streit90} uses a detection step and~\cite{Barret93} uses an approximation.

\item Lastly, the tracking method is entirely unsupervised, with a maximum likelihood 
hyperparameter estimation. This is not a straightforward task in the context of frequency
tracking, since the non-linear character of the data as functions of frequencies prevents
the explicit handling of the likelihood function of the hyperparameter. We have developed
an EM-like gradient procedure, inspired by~\cite{Levinson82,Lange95,McLachlan97}. It can be derived
only after discretizing the frequencies on a finite grid. 

\een

The paper is organized as follows. The notation, signal model and assumptions are defined
in Section~\ref{Statement}. Section~\ref{Proposed} contains the proposed regularized method
and Section~\ref{Discrete} gives a discrete approximation. Section~\ref{Hyper} is devoted
to the estimation of hyperparameters. The performance of the proposed method is
demonstrated by the computer simulations in Section~\ref{Results}, while
Section~\ref{Conclus} gives our conclusion and describes possible extensions.

\section{Statement, notations and assumptions}\label{Statement}

In Doppler imaging, the signals to be analyzed occur as a set of complex signals
$\Yc=[\yb_1, \dots, \yb_T]$ juxtaposed spatially, in $T$ range
bins~\cite{Talhami88,Barton97}. The data record $\yb_t=[y_t(1), \dots, y_t(N)]\T$, (``t''
denotes the matrix transpose) is extracted from a cisoid in additive complex noise. The
amplitude and the frequency of the cisoid are $a_t\in\eC$ and $\nu_t\in\eR$:
\beq \label{ModelObserv}
\yb_t	= a_t \, \zb(\nu_t) + \bb_t 
		= a_t \, [1, \dots, e^{j 2 \pi \nu_t (N-1)} ]\T + \bb_t \,.
\eeq
The vectors $\nub = [\nu_1,\dots,\nu_T]\T$ and $\ab = [a_1,\dots,a_T]\T$ collect the
frequencies and corresponding amplitudes. Finally, the true parameters are denoted with a
star. This paper builds a robust estimate $\wh{\nub}$ for $\nub^\star$ on the basis of data
set $\Yc$ (see Fig.~\ref{Data} for a simulated example). 

\begin{figure}[ht]\setlength{\tabcolsep}{2pt}
\bcc
\btabu{cc}
\rotatebox{90}{\small \hspace{0.75cm} Frequency $\nu$ \hspace{1.5cm} Sample \# $n$}
&\includegraphics[height=6cm]{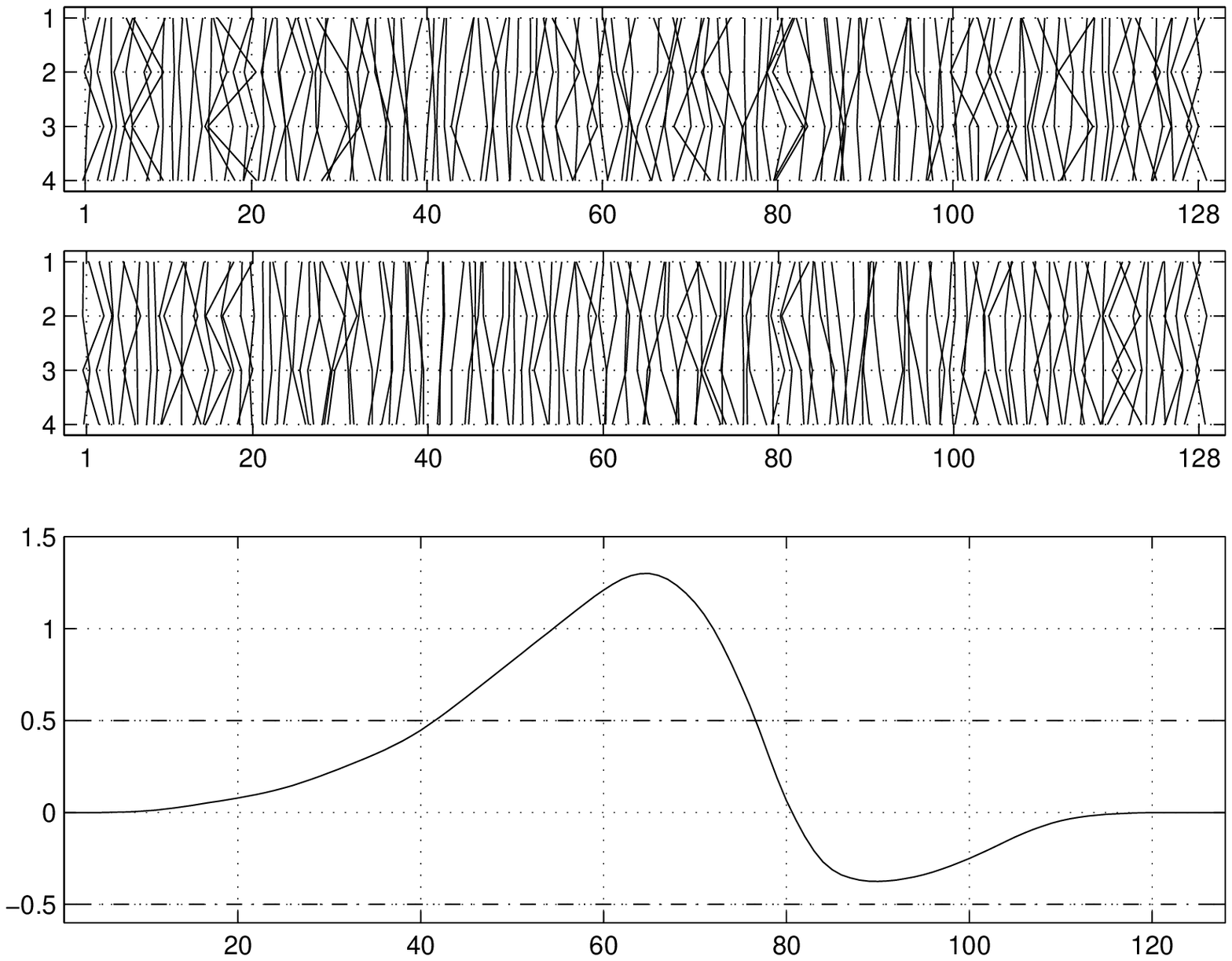}\\
& \small Depth $t$
\etabu
\ecc
\caption[Observed data and true frequency]{Simulated observations over $T=128$ range bins
with $N=4$ samples per bin. From top to bottom: real parts, imaginary parts of the data
$\yb_t$, and the true frequency sequence $\nu_t^\star$.}
\label{Data}\label{Vrai}
\end{figure}

\begin{remark}---
Model~(\ref{ModelObserv}) is frequently used for spectral problems; it has three main
features. First, while it is linear \wrt $a_t$, it is not so \wrt $\nu_t$: the problem to
be solved is non-linear. Second, $\zb(\nu_t)$ is a 1-periodic function \wrt $\nu_t$ and
this causes the difficulties of aliasing, frequency ambiguity, likelihood periodicity, \etc
Lastly, this periodicity is also the keystone of the paper: aliasing is inverted using a
coherent statistical approach that takes periodicity in consideration.
\end{remark}

The following definition of periodicity is used throughout the paper.
\begin{definition}--- \label{DefPeriodic}
Let $A\subset \eR^T$ and $\varphi~: A \rightarrow \eR$. Let us note $\unb=[1,\dots,1]\T \in
\eR^T$. $\varphi$ is said:
\bit
\item \emph{separately-1-periodic} (S1P) if $\forall \ub \in A$, $\forall \kb \in \eZ^T$
(such that $\ub+\kb \in A$): $\varphi(\ub) = \varphi(\ub + \kb)$.
\item \emph{globally-1-periodic} (G1P) if $\forall \ub \in \eR^T$, $\forall k_0 \in \eZ$
(such that $\ub+k_0 \unb \in A$): $\varphi(\ub) = \varphi(\ub + k_0 \unb)$.
\eit
\end{definition}

The proposed estimation method deals with periodicity and aliasing inversion thanks to the
following assumptions. They are stated for the sake of simplicity and calculation
tractability as well as coherence with the applications under the scope of this paper.
\bit

\item Parameter dependence. 

	\bit
	\item $\HD_1$: $\ab$, $\nub$ and the $\bb_t$ are independent
	\eit

\item Law for measurement and modeling noise $\bb_t$.

	\bit
	\item $\HD_2^\aD$: each $\bb_t$ is $\Nc(r_b I_N)$
	\item $\HD_2^\bD$: the sequence of $\bb_t$ is itself white
	\eit

\item Law for parameters $\ab$ and $\nub$.

	\bit
	\item $\HD_3^\aD$: $\ab$ is $\Nc(r_a I_T)$, \ie white
	\item $\HD_3^\bD$: $\nub$ is, on the contrary, correlated: $\Nc(R_\nu)$
	\eit

\eit
where $\Nc(R)$ stands for a complex zero-mean Gaussian vector with covariance $R$, and
$I_P, P\in\eN^*$ denotes the $P\times P$ identity matrix. 

The first assumption $\HD_1$ is quite natural since no information is available about the
relative fluctuations of noise and objects. The assumptions $\HD_2^\aD$, and $\HD_2^\bD$
are also natural since no correlation structure is expected in noise. Similarly, we have no
information about the variation of the amplitude sequence, so an independent law is used. A
Gaussian law is preferred ($\HD_3^\aD$) to make the calculations tractable. Contrarily, the
smoothness of the frequency sequence is modeled as a positive correlation. A Markovian
structure (specified below) is a simple, useful way to account for it. Several choices are
available, but the Gaussian one is also stated for the sake of simplicity ($\HD_3^\bD$).

\section{Proposed method}\label{Proposed}

\subsection{Likelihood}\label{ProposedLikelihood}

Assumption $\HD_2^\aD$ yields a parametric structure for each likelihood function $f(\yb_t
\I \nu_t, a_t)$:
\beqx
f(\yb_t \I \nu_t, a_t) = (\pi r_b)^{-N} \Exp{-\frac{1}{r_b} CLL(\nu_t, a_t)}
\eeqx
involving the opposite of the logarithm of the likelihood function (up to constant terms)
\ie the Co-Log-Likelihood (CLL):
\beqx
CLL(\nu_t, a_t) = \cro{\yb_t - a_t \zb(\nu_t)}^{\dag} \cro{\yb_t - a_t \zb(\nu_t)}  \,. 
\eeqx
From a deterministic standpoint, $CLL(\nu_t, a_t)$ is clearly the Least Squares (LS)
estimation criterion. 

Considering the whole frequency vector $\nub$ and the whole data set $\Yc$, assumption
$\HD_2^\bD$ yields:
\beq\label{VraisJoint}
 f(\Yc \I \nub, \ab)=(\pi r_b)^{-NT} \Exp{- \frac{1}{r_b} CLL(\nub, \ab)} 
\eeq
where the global CLL is a global LS criterion:
\beqx
CLL(\nub, \ab) = \sum_{t=1}^T CLL(\nu_t, a_t) \,.
\eeqx

\begin{remark}--- \label{PeriodicLikelihood}
According to Definition~\ref{DefPeriodic}, the likelihood function $CLL(\,\cdot\,,\ab)$ is
S1P for all $\ab\in\eC^T$. So, two configurations $\nub$ and $\nub+\kb$ ($\kb\in\eZ^T$) for
the frequency sequence are \emph{equi-likelihood}. As a consequence, an ML approach suffers
from $T$ independent frequency ambiguities.
\end{remark}

\subsection{Amplitude law and marginalization}\label{AmpliMarginalize}

The parameters of interest are the frequencies, while the amplitudes are nuisance
parameters. These are integrated out of the problem in the usual Bayesian approach.

Given separability assumption $\HD_1$ one has $f(\nub,\ab)=f(\nub)f(\ab)$ and the marginal
law can easily be deduced:
\beqx
f(\Yc, \nub) = f(\nub) \int_{\ab} f(\Yc \I \ab,\nub) f(\ab) d\ab = f(\nub) f(\Yc \I \nub) \,.
\eeqx
The joint law for the amplitudes is separable according to assumption $\HD_3^\aD$. Since
likelihood~(\ref{VraisJoint}) is also separable, marginalization can be performed
independently.  
\beq \label{EqAmpliMarginalize}
f(\Yc \I \nub)	= \prod_{t=1}^T \int_{a_t} f(\yb_t \I \nu_t , a_t) f(a_t) da_t.
					= \prod_{t=1}^T f(\yb_t \I \nu_t). 
\eeq

The Gaussian amplitude assumption $\HD_3^\aD$ results in analytic derivations and yield the
marginal likelihood for the data $\yb_t$ given $\nu_t$: a zero mean Gaussian vector. Its
covariance $R_t$ is given in Appendix~\ref{ObservCondFreq} as well as its
determinant~(\ref{DetR}) and its inverse~(\ref{InvR}). $f(\yb_t \I \nu_t)$ then reads:
\beqn \label{LikelihoodUnNu}
f(\yb_t \I \nu_t) = \beta \, \Exp{-\gamma_t} \, \Exp{\alpha P_t(\nu_t)} 
\eeqn
with $\alpha=\froc{N r_a}{(r_b(N r_a  + r_b))}$, $\beta=\froc{\pi^{-N} r_b^{1-N}}{(N r_a +
r_b)}$, $\gamma_t=\yb_t^{\dag} \yb_t/r_b$, and $P_t$ is the periodogram of vector $\yb_t$
\beqx
P_t(\nu_t) = \frac{1}{N} \left| \sum_{n=1}^N y_t(n) e^{-2j\pi \nu_t n} \right|^2 \,.
\eeqx

The joint law for the whole data set given the frequency sequence is obtained by the
product~(\ref{EqAmpliMarginalize}):
\beq \label{LikelihoodNu}
f(\Yc \I \nub) = \beta^T \Exp{-\gamma} \Exp{-\alpha CLML(\nub)}
\eeq
where $\gamma$ is the sum of the $\gamma_t$ for $t\in\eN^*_T=\left\{1,\dots,T\right\}$ and
where CLML is the Co-Log-Marginal-Likelihood  
\beq \label{CritLikelihoodNu}
CLML(\nub) = -\sum_{t=1}^T P_t(\nu_t)
\eeq
the opposite of the sum of the periodograms of data $\yb_t$ at frequency $\nu_t$, in gate
$t$. 

\begin{remark}--- \label{PeriodicMarginalLikelihood}\label{ArgMaxPer}
This remark is the marginal counterpart of Remark~\ref{PeriodicLikelihood}. As well as
$CLL(\,\cdot\,,\ab)$, $CLML(\,\cdot\,)$ is S1P: there are still many ambiguities as in the
non-marginal case. This was expected since no information about the frequency sequence has
been accounted for in $CLML(\nub)$ \wrt $CLL(\nub,\ab)$. In contrast, periodicity will be
eliminated in the next subsection by accounting for the frequency sequence smoothness.
\end{remark}

\subsection{\Prior law for frequency sequence}

Unlike amplitudes, the frequency sequence is smooth. A Markovian structure accurately
accounts for this information, and there are many algorithms suited to computing this
structure. The choice of the family law is not crucial for using these algorithms, but we
have used the Gaussian family:
\beqx
f(\nu_{t+1} \I \nu_{t}) = (2 \pi r_\nu)^{-1/2} 
									\Exp{- \frac{1}{2r_\nu} (\nu_{t+1} - \nu_t)^2} \,. 
\eeqx
The complete law for the chain also involves the initial state. It is assumed to be
uniformly distributed over a symmetric set $\Kbb$ defined by $K\in\eN^*$: $\Kbb=[-K/2;
+K/2]$. So $f(\nu_1)=(1/K) \unbb_K^0(\nu_1)$ where $\unbb_K^0$ is 1 in $\Kbb$ and 0
outside. 

The recursive conditioning rule immediately yields: 
\beqn \label{PriorNu}
f(\nub) = (2 \pi r_\nu)^{-(T-1)/2} \Exp{- \frac{1}{2r_\nu} CLP(\nub)} \,,
\eeqn
where $CLP(\nub)$ is the Co-Log-Prior:
\beq \label{CritPriorNu}
CLP(\nub) =  \widetilde{K} \unbb_K^\infty (\nu_1)  + \sum_{t=1}^{T-1} (\nu_{t+1} - \nu_t)^2 \,,
\eeq
$\widetilde{K}=2r_\nu\log K$ and $\unbb_K^\infty$ is 1 in $\Kbb$ and $+\infty$ outside. In the
deterministic framework $CLP(\nub)$ is a quadratic norm for the first-order differences,
namely a regularization term~\cite{Demoment89,Tikhonov77,Hunt77}.

\subsection{\Post law}

Fusion of \prior-based and data-based information is achieved by the Bayes rule, which
provides the \apost density for $\nub$:
\beqx
f(\nub \I \Yc)  = \froc{f(\Yc \I \nub) f(\nub)}{f(\Yc)} \,.
\eeqx
The marginal law $f(\Yc)$ for the whole data set $\Yc$ is not analytically tractable,
essentially due to the non-linearity of the periodogram \wrt $\nu_t$ and the correlated
structure of $\nub$. Fortunately, this \pdf does not depend on $\nub$,  so the \apost density
remains explicit up to a positive constant.  \Prior structure of
Eq.~(\ref{PriorNu}-\ref{CritPriorNu}) and likelihood structure of
Eq.~(\ref{LikelihoodNu}-\ref{CritLikelihoodNu}) immediately yield the \post law:
\beq \label{PostLaw}
 f(\nub \I \Yc)  \propto \Exp{-\alpha CLPL(\nub)}
\eeq
where the Co-Log-Posterior-Likelihood function (CLPL) reads:
\beq \label{CritPostLaw}
CLPL(\nub) = -\sum_{t=1}^T P_t(\nu_t) + \lambda \sum_{t=1}^{T-1} (\nu_{t+1} - \nu_t)^2  
				+ \unbb_K^\infty (\nu_1)
\eeq
with $\lambda=1/2\alpha r_\nu$, up to irrelevant constants. In the deterministic framework,
$CLPL$ is a Regularized Least Squares (RLS) criterion. It has three terms, one measures
fidelity to the data, the second measures fidelity to the \prior smoothness and the third
enforces the first frequency $\nu_1\in\Kbb$. The regularization parameter $\lambda$
(depending on hyperparameters $\rb=[r_a,r_b,r_\nu]$) balances the compromise between
\prior-based and data-based information.

\subsection{Point estimate}

As a point estimate, a popular choice is the Maximum \textit{A Posteriori} (MAP) \ie the
maximizer of the \post law of Eq.~(\ref{PostLaw}) or the minimizer of the RLS
criterion~(\ref{CritPostLaw}): 
\beq \label{MAP}
\wh{\nub}\MAP = \argmax_{\nub\in\eR} f(\nub \I \Yc) = \argmin_{\nub\in\eR} CLPL(\nub) \,.
\eeq

\begin{figure}[ht]\setlength{\tabcolsep}{2pt}
\bcc
\btabu{cc}
\rotatebox{90}{\small \hspace{0.5cm} $CLPL$ \hspace{1.25cm}$CLP$ \hspace{1cm} $CLML$}
&\includegraphics[height=6cm]{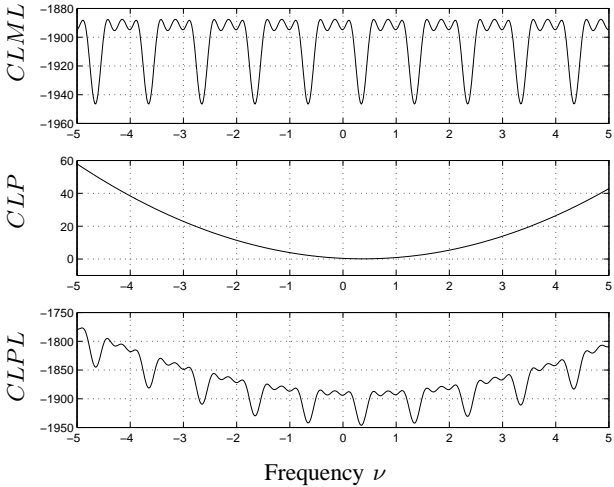} \\
&\small Frequency $\nu$
\etabu
\ecc
\caption[Criteria]{Typical form of criteria. From top to bottom: $CLML(\nub)$ (periodic),
$CLP(\nub)$ (quadratic) and $CLPL(\nub)$ as a function of $\nu_t$ ($t=50$).
Regularization breaks periodicity.}
\label{CritFreq}
\end{figure}

\begin{remark}--- \label{NonPeriodic}
This remark is the posterior counterpart of Remarks~\ref{PeriodicLikelihood}
and~\ref{PeriodicMarginalLikelihood}. Whereas CLL and CLML are S1P, CLPL is not:
regularization breaks periodicities, favors solutions according to \prior probabilities,
and enables some ambiguities to be removed. Nevertheless, a global indetermination remains:
$CLPL$ is a G1P function. This is essentially due to ($i$) the marginal likelihood  CLML is
a S1P function and ($ii$) the regularization term CLP is a G1P function (since it only
involves frequency differences). As a consequence, two frequency profiles, different from a
constant integer level remain \emph{equi-likelihood}. Finally, the latter indeterminacy can
be removed by  choosing an appropriate $K$: $K=1$ enforces the first frequency $\nu_1$ to
remain in $(-0.5,+0.5]$ and the corresponding $CLPL$ is no longer G1P.
\end{remark}

\medskip\begin{proposition}--- \label{PropDiffMAP}
With the previous notations and definitions, the MAP estimate is such that:
\beq \label{EqPropDiffMAP}
| \wh{\nu}\MAP_{t+1} -\wh{\nu}\MAP_{t} | \leqslant 1/2 ~~~\FOR~~ t\in\eN_{T-1}^*
\eeq
\end{proposition}

\medskip\begin{proof}
See appendix~\ref{DemPropDiffMAP}.
\end{proof}\medskip

\subsection{Optimization stage}\label{GradEtHessien}

The proposed approach allows ambiguous periodicity to be removed at the expense of
accepting local minima in the built energy~(\ref{CritPostLaw}). A gradient
procedure~\cite{Bertsekas95} can achieve \emph{local} minimization of~(\ref{CritPostLaw})
and $CLPL$ gradient involves the periodograms derivatives
\beqx
P_t'(\nu_t) = 2j\pi \sum_{n=1-N}^{N-1} n \wh{c}_t(n) e^{2j\pi \nu_t n}
\eeqx
when rewriting $P_t(\nu_t)$ as a function of empirical correlation lags $\wh{c}_t(n)$ of
the signal $\yb_t$. It is also possible to calculate the second-order derivative
\beqx
P_t''(\nu_t) = -4\pi^2 \sum_{n=1-N}^{N-1} n^2 \wh{c}_t(n) e^{2j\pi \nu_t n}
\eeqx
and to implement second-order descent algorithms. 

There are several ways of coping with \textit{global} optimization, \eg graduated
non-convexity~\cite{Blake87,Nikolova98}, stochastic algorithms such as simulated
annealing~\cite{Geman84,Robert96}. We have used a dynamic programming procedure for
computational simplicity. It is based on a discrete approximation of the \prior law for the
frequencies. This approximation allows global optimization (on an arbitrary fine discrete
frequency grid) and provides a convenient framework for estimating hyperparameters.

\section{Discrete state Markov chain}\label{Discrete}

This section is devoted to a discrete approximation to
\ben
\item maximize \post law for the frequency sequence $\nub$
\item build an ML procedure for estimating hyperparameters. 
\een
We have therefore introduced an equally spaced discretization of the frequency range
$[\nu_\mD;\nu_\MD]$ in $P$ states $\nu^1, \dots, \nu^P$ ($\nu_\MD=-\nu_\mD=2.5$ and $P=128$
in our simulations).

\subsection{Probabilities}

Discretization and normalization of the \aprio law~(\ref{PriorNu}) yields the state
transition probabilities:
\beqn
\Pbb_t(p,q)	&=& \Prob{\nu_{t+1} = \nu^p \I \nu_t = \nu^q} \nonumber\\
				&=& \frac{\exp \left( - (\nu^p- \nu^q)^2 /2r_\nu\right)}
            {\sum_{p=1}^{P} \exp \left(- (\nu^p- \nu^q)^2/ 2r_\nu\right)} \,.\label{ProbaTransit}
\eeqn
Note that $\Pbb_t$ does not depend upon $t$, \ie the proposed chain is homogeneous
$\Pbb_t=\Pbb$. The full state model also includes the initial probabilities $\pbb(p)$, chosen  
constant over $(-0.5,+0.5]$ (see Remark~\ref{NonPeriodic}).

The marginal (\wrt amplitudes) likelihood function for the observation sequence given by
Eq.~(\ref{LikelihoodUnNu}) yields the observation probability distribution
$\Obb_t(p)=f(\yb_t \I \nu_t=\nu^p)$.

\subsection{Available algorithms}

The Markov chain is now convenient for using algorithms given
in~\cite{Rabiner86,Forney73},  the Viterbi and the Forward-Backward algorithms. They enable
us to compute
\bit
\item the MAP and 
\item the hyperparameters likelihood as well as its gradient.
\eit

\subsubsection{The Viterbi algorithm} \label{ViterbiAlgo}

The Viterbi algorithm, Appendix~\ref{AnnexViterbi}, has been implemented to cope with
global optimization (on a discrete grid), and performs a step-by-step optimization of the
\post law. The required observation probabilities are also readily pre-computable by FFT.

\subsubsection{Forward~--~Backward algorithm}

We have used a normalized version of the procedure, as recommended
in~\cite{Devijver88,Devijver85} to avoid computational problems. It is  founded on forward
and backward probabilities:
\beqx 
\Fc_t(p)=\frac{\Prob{\Yc_1\T , \nu_t=\nu^p}}{\Prob{\Yc_1\T}}    ~\AND~
\Bc_t(p)=\frac{\Prob{\Yc_{t+1}^T \I \nu_t=\nu^p}}{\Prob{\Yc_{t+1}^T \I \Yc_1\T}}
\,,
\eeqx
where $\Yc_t^{t'} = [\yb_t, \dots, \yb_{t'} ]$ denotes the partial observation matrix from
time $t$ to $t'$.

The (count-up) Forward algorithm, given in Appendix~\ref{AnnexForward}, computes non
normalized probabilities $\overline{\Fc}_t(p)$, normalization coefficients $\Nc_t$ and the
$\Fc_t(p)$ themselves. As a result, the observation likelihood can be deduced:
\beq \label{ProbaData}
\Prob{\Yc}  = \prod_{t=1}^P  \Nc_t \,.
\eeq
It is useful for estimating ML hyperparameters in Section~\ref{Hyper}. The (count-down) Backward
step, described in Appendix~\ref{AnnexBackward}, yields marginal \apost probabilities
(see~\cite[p.10]{Rabiner86}):
\beq \label{MargPost}
p_t(p) = \Prob{\nu_t = \nu^p \I \Yc} = \Fc_t(p) \Bc_t(p)        
\eeq
and double marginal \apost probabilities (see~\cite[p.11]{Rabiner86}) 
\beqn
p_t(p,q)	&=& \Prob{\nu_{t-1}=\nu^q, \nu_t = \nu^p \I \Yc}\nonumber\\
			&=& \Nc_t \, \Fc_{t-1}(p) \, \Bc_t(q) \, \Pbb(p,q) \, \Obb_t(q) \,, \label{DoubleMargPost}
\eeqn
both needed to calculate the likelihood gradient.

\section{Estimating Hyperparameters}\label{Hyper}

The MAP estimate of Eq.~(\ref{MAP}) depends upon a unique regularization parameter
$\lambda$, function of three hyperparameters $\rb=[r_a,r_b,r_\nu]$. This section is devoted
to their estimation using the available data set $\Yc$. 

Estimating hyperparameters within the regularization framework is generally a delicate
problem. It has been extensively studied, several techniques have been proposed and
compared \cite{Golub79,Titterington85,Hall87,Thompson91,Fortier92,Giovannelli96} and the
preferred strategy is founded on ML.

The ML estimation  consists of \textit{(i)} expressing the Hyperparameter Likelihood (HL) as
$HL_{\Yc} (\rb)=f(\Yc)$  and \textit{(ii)} maximizing the resulting function.  Although we
have chosen a simple Gaussian law, \nub cannot be marginalized in closed form because \nub
enters $f(\Yc|\nub)$ in a complex manner.  Fortunately, the discrete state approximation of
Section~\ref{Discrete} provides a satisfactory solution to this problem. It also allows us to
devise several kinds of algorithms for local maximization of the likelihood.  One
such scheme is the acknowledged EM (Expectation-Maximization) algorithm, although its
application reveals uneasy in the present context of a parametric model of hidden Markov
chain (\cite{Levinson82} provides a meaningful discussion of such situations, see
also~\cite{Lange95,McLachlan97}). Subsection~\ref{gradient} is devoted to the EM framework, within which
a gradient procedure is proposed. Subsection~\ref{likelihood} deals with the computation of
the likelihood and proposes a simple coordinatewise descent procedure.

\subsection{Hyperparameter likelihood}\label{likelihood}

The hyperparameter likelihood $HL_{\Yc}$ can be deduced from the joint law for ($\nub,\Yc$)
by frequency marginalization: 
\beqx
HL_{\Yc} (\rb)  = \dispsty\sum_{p_1,\dots,p_T = 1}^P 
						\Prob{\Yc , \nu_1=\nu^{p_1},\dots,\nu_T=\nu^{p_T}}
\eeqx
but the indices run over $P^T$ states, so the above summation is not directly tractable.
However, the Forward procedure efficiently achieves a recursive marginalization: it yields
$HL_{\Yc}(\rb)$ according to Eq.~(\ref{ProbaData}) and requires about $TP^2$ calculations.

Let us introduce the Co-Log-HL (CLHL) to be minimized \wrt hyperparameters vector $\rb$:
\beqnx \label{HyperMV}
\wh{\rb}\ML = \argmin_{\rb} CLHL_{\Yc}(\rb).
\eeqnx
One possible optimization scheme is a coordinatewise descent algorithm, with a golden
section line search~\cite{Bertsekas95}. But a more efficient scheme may be a gradient 
algorithm~\cite{Bertsekas95}.

\subsection{Likelihood gradient}\label{gradient}

The EM algorithm relies on an auxiliary function, usually denoted
$Q$~\cite{Baum70,Liporace82} built on two hyperparameter vectors $\rb$ and $\rb'$ by
completing the observed data set $\Yc$ with parameters to be marginalized $\nub$:
\beqnx
Q(\rb, \rb')	&=& E_{\nub} \Big[ \log( \Prob{\nub,\Yc ; \rb'}) \,\Big|\, \Yc ; \rb \Big] \\
					&=& \sum_{\nub} \log \Prob{\nub, \Yc ; \rb'} ~ \Prob{(\nub \I \Yc ; \rb)}. 
\eeqnx
With the proposed notations, usual hidden Markov chains calculations yield:
\beqn \label{QDeveloppe}
Q(\rb, \rb')&=& \sum_{t=2}^{T} \sum_{p,q=1}^P p_t(p,q) \, \log\Pbb'(p,q)\\
				&+&  \sum_{p=1}^P \pbb(p) \, \log\pbb'(p) \,
						+ \sum_{t=1}^T \sum_{p=1}^P p_t(p) \, \log\Obb_t'(p)\nonumber
\eeqn
where:
\bit
\item $(\pbb',\Pbb',\Obb')$  and $(\pbb,\Pbb,\Obb)$ are parameters of the model under
hyperparameters $\rb'$ and $\rb$, respectively. 
\item $p_t(p)$ and $p_t(p,q)$ denote the \apost marginal laws defined by~(\ref{MargPost})
and~(\ref{DoubleMargPost}), under hyperparameters $\rb$. 
\eit

The $k$th iteration of the EM scheme maximizes $Q(\rb^{(k-1)}, \rb')$ as a function of
$\rb'$, to yield $\rb^{(k)}$ as the maximizer. Unfortunately, it seems impossible to derive
an explicit expression for such a maximizer. However, an alternate route can be followed,
given the key property:
\beqx
\frac{\partial Q(\rb, \rb')}{\partial \rb'} \Big|_{\rb' = \rb}
		= - \frac{\partial CLHL_{\Yc}(\rb)}{\partial \rb}.
\eeqx
As suggested by~\cite{Levinson82}, this property enables us to calculate the gradient of
$CLHL_{\Yc}(\rb)$ as the derivative of~(\ref{QDeveloppe})~:
\beqn
\frac{\partial Q}{\partial r_a'}
 &=& \sum_{t=1}^T \sum_{p=1}^P p_t(p) \frac{\partial \log\Obb_t'(p)}{\partial r_a'}
\label{SumDuGradDebut}\\
\frac{\partial Q}{\partial r_b'}
 &=& \sum_{t=1}^T \sum_{p=1}^P  p_t(p)\frac{\partial \log\Obb_t'(p)}{\partial r_b'} \\
\frac{\partial Q}{\partial r_\nu'}
 &=& \sum_{t=2}^T \sum_{p,q=1}^P p_t(p,q) \frac{\partial \log\Pbb'(p,q)}{\partial r_\nu'}.        \label{SumDuGradFin}
\eeqn
The encountered derivatives $\froc{\partial \log\Obb'(p)}{\partial r_a'}$,
$\froc{\partial \log\Obb'(p)}{\partial r_b'}$ and $\froc{\partial \log\Pbb'(p,q)}{\partial
r_\nu'}$ respectively read:
\beqnx
&&\frac{-N}{N r_a' + r_b'} + \frac{N}{(N r_a' + r_b')^2} P_t(\nu^p) \\
&& \frac{1-N}{r_b'} - \frac{1}{N r_a' + r_b'} + \frac{\yb_t^{\dag} \yb_t}{r_b^{'2}}  -
	\frac{N r_a'(N r_a' + 2r_b')}{r_b^{'2}(N r_a' + r_b')^2} P_t(\nu^p) \\
&& \frac{1}{2 r_\nu^{'2}} \Big( (\nu^q - \nu^p )^2 
	- \sum_{r=1}^P (\nu^r - \nu^p)^2 \Pbb'(p,r) \Big)
\eeqnx
by derivation of~(\ref{LikelihoodUnNu}) and~(\ref{ProbaTransit}). Finally the likelihood
gradient is readily calculated and a gradient procedure can be applied.

\begin{figure*}[ht]\setlength{\tabcolsep}{2pt}
\bcc
\btabu{cccccc}
\rotatebox{-90}{\small \hspace{1.5cm}$\log_{10} r_b$}&
\includegraphics[height=4cm,width=4cm,bb=400 695 222 505,angle=-90,clip]{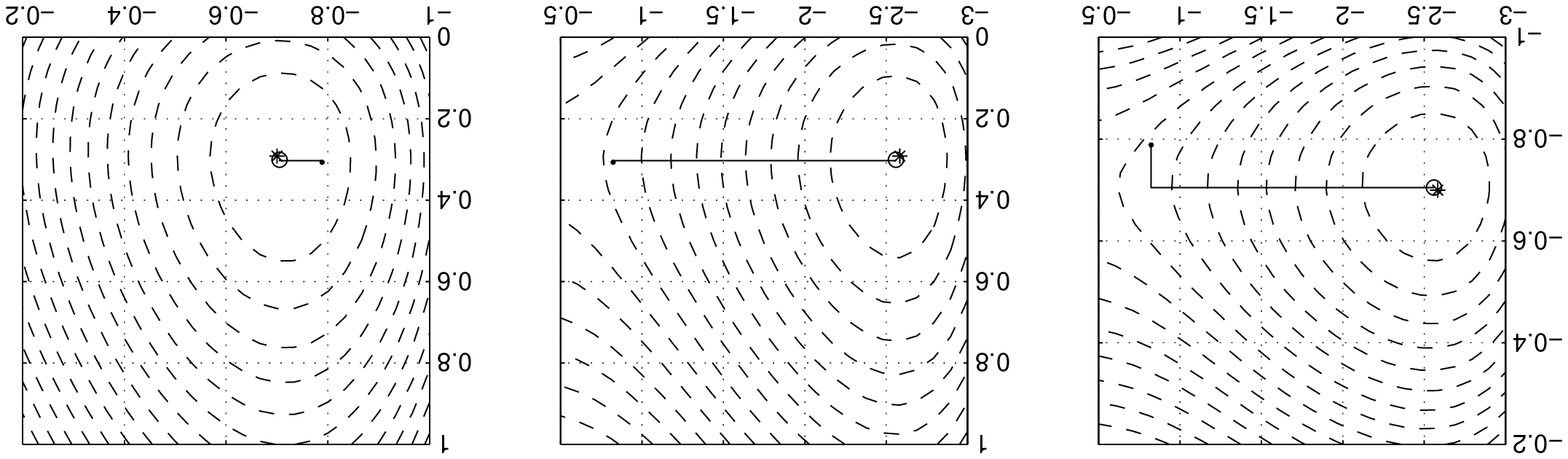}&
\rotatebox{-90}{\small \hspace{1.5cm}$\log_{10} r_a$}&
\includegraphics[height=4cm,width=4cm,bb=400 480 222 294,angle=-90,clip]{fig3a}&
\rotatebox{-90}{\small \hspace{1.5cm}$\log_{10} r_a$}&
\includegraphics[height=4cm,width=4cm,bb=400 270 222 80,angle=-90,clip]{fig3a}\\
 & \small $\log_{10} r_\nu$ & & \small $\log_{10} r_\nu$ & & \small $\log_{10} r_b$ \\
\rotatebox{-90}{\small \hspace{1.5cm}$\log_{10} r_b$}&
\includegraphics[height=4cm,width=4cm,bb=400 695 222 505,angle=-90,clip]{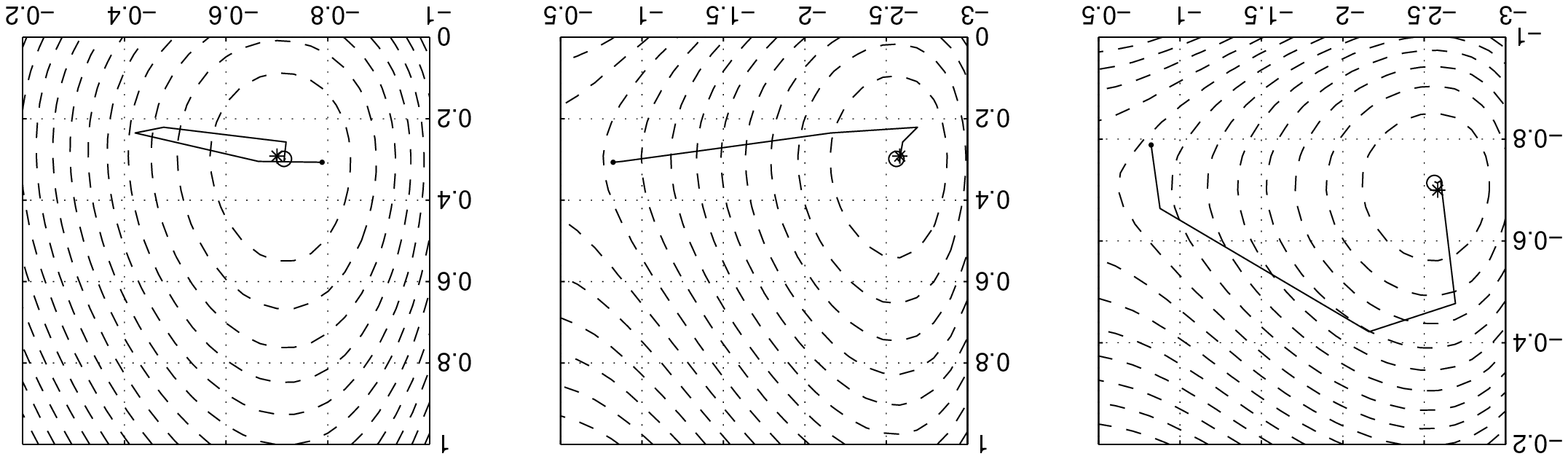}&
\rotatebox{-90}{\small \hspace{1.5cm}$\log_{10} r_a$}&
\includegraphics[height=4cm,width=4cm,bb=400 480 222 294,angle=-90,clip]{fig3b}&
\rotatebox{-90}{\small \hspace{1.5cm}$\log_{10} r_a$}&
\includegraphics[height=4cm,width=4cm,bb=400 270 222 80,angle=-90,clip]{fig3b}\\
 & \small $\log_{10} r_\nu$ & & \small $\log_{10} r_\nu$ & & \small $\log_{10} r_b$\\ 
\etabu
\ecc
\caption[Hyperparameter likelihood]{Hyperparameter likelihood: typical behavior. Level sets
of $CLHL$ are plotted as dashed lines (-\,-). The minima are located by a star ($*$),
starting points (empirical estimates) by a dot ({\LARGE .}) and final estimate by a circle
(o). First row gives coordinate-wise algorithm and second row gives gradient algorithm.
First column: $CLHL(\wh{r}_a\ML,r_b,r_\nu)$, second column: $CLHL(r_a,\wh{r}_b\ML,r_\nu)$,
third column: $CLHL(r_a,r_b,\wh{r}_\nu\ML)$. Each figure is $\log_{10}$-scaled.}
\label{FigVraisHyper}
\end{figure*}

\section{Simulation results and comparisons}\label{Results}

The previous Sections introduced a regularized method for frequency tracking and estimating
hyperparameters. This Section demonstrates the practical effectiveness of the proposed
approach by processing\footnote{Algorithms have been implemented using the computing
environment \textit{Matlab} on a Personal Computer, Pentium~III, with a 450~MHz CPU and 128
MB of RAM.} simulated signals shown in Fig.~\ref{Data}.  

\subsection{Hyperparameter estimation}\label{ResultsHyper}

The hyperparameter likelihood function $CLHL$ was first computed on a fine discrete grid of
$25\times 25\times 25$ values resulting in the level sets shown in
Fig.~\ref{FigVraisHyper}. The function is fairly regular, and has a single minimum.

\begin{table*}[htb] 
\begin{center}
\begin{tabular}{|c|c|c|c|c|c|c|} \hline
Method  & Reached minimum       &$\log_{10}\wh{r}\ML_a$& $\log_{10}\wh{r}\ML_b$&
$\log_{10}\wh{r}\ML_\nu$ & Grad./Fun. &Time ($s$)\\
\hline\hline
(1a)            & 4.513 $10^{2}$        & 0.297   & -0.685 & -2.424        & 17/59 & 5.55  \\
(1b)            & 4.495 $10^{2}$        & 0.297   & -0.679 & -2.519        & 13/87 & 5.92  \\ 
(2a)            & 4.494 $10^{2}$        & 0.292   & -0.678 & -2.537        & 9~/49 & 3.77  \\
(2b)            & 4.494 $10^{2}$        & 0.299   & -0.681 & -2.554        & 13/92 & 6.14  \\
(3a)            & 4.498 $10^{2}$        & 0.297   & -0.695 & -2.589        & 9~/53 & 4.07  \\
(3b)            & 4.494 $10^{2}$        & 0.298   & -0.679 & -2.547        & 13/92 & 6.21  \\
(4a)            & 4.497 $10^{2}$        & 0.283   & -0.674 & -2.507        & 7~/40 & 3.12  \\
(4b)            & 4.500 $10^{2}$        & 0.297   & -0.685 & -2.618        & 9~/75 & 4.84  \\\hline
(5)             & 4.495 $10^{2}$        & 0.300   & -0.671 & -2.559        & 0~/81 & 3.41  \\\hline
\end{tabular}
\end{center}
\caption[Descent algorithm comparison]{Descent algorithm comparison. The first column
gives the method at work: (1) usual gradient, (2) Vignes correction, (3) bisector
correction and (4) Polak-Ribi\`ere pseudo-conjugate direction. (\textnormal{a}) no
interpolation and (\textnormal{b}) quadratic interpolation. (5) is the coordinate-wise
descent method. The following columns show the reached minimum and the minimizer. The
sixth column gives the number of gradients and function calculus while the last gives
computation times in seconds ($s$).}
\label{TableHyperTemps}
\end{table*}
        
The hyperparameters are tuned using two classes of descent algorithms:
\bit
\item a coordinate-wise descent algorithm
\item a gradient descent algorithm.
\eit
The latter employs several descent directions: usual gradient, bisector correction, Vignes
correction and Polak-Ribi\`ere pseudo-conjugate direction. Two line search methods have
also been implemented: usual dichotomy and quadratic interpolation. The starting point
remains the empirical hyperparameter vector described in Appendix~\ref{HyperEmpiric}. 

All the strategies provide the correct minimizer and they are compared
Table~\ref{TableHyperTemps} and Fig.~\ref{FigVraisHyper}. The usual gradient generated
zig-zagging trajectories and was slower than the other strategies. The three corrected
direction strategies were 25 to 40 \% faster than the uncorrected ones, with the
Polak-Ribi\`ere pseudo-conjugate direction having a slight advantage. In contrast,
interpolation did not result in any improvement within the corrected direction class. 

The coordinate-wise descent algorithm performed well, since it does not require any gradient
calculation. Gradient calculus needs much more computation than the likelihood itself, due
to summations in Eq.~(\ref{SumDuGradDebut})-(\ref{SumDuGradFin}). Likelihood calculus took
$0.05~s$, while gradient calculus required $0.2~s.$, \ie about four times more. 

We have therefore adopted the two fastest methods: coordinate-wise and Polak-Ribi\`ere
pseudo-conjugate gradient, which took less than $3.5~s$. Fig.~\ref{FigVraisHyper} also
illustrates the convergence.

\subsection{Frequency tracking}

The optimization procedure used to compute the MAP (given ML hyperparameters) consisted of
applying the Viterbi algorithm (described in  Section~\ref{ViterbiAlgo}). The solution was
used as the starting point for the gradient or the Hessian procedure (described in
Section~\ref{GradEtHessien}). The Viterbi algorithm explored the whole set of possible
frequencies (on a discrete grid) and found the correct interval for each frequency, while
the gradient or Hessian procedure locally refined the optimum. Table~\ref{TableFreqEstTps}
shows the computation times. We adopted the Hessian procedure, since it performed almost 10
times faster. 

\begin{table}[hbt] 
\begin{center}
\begin{tabular}{|c|c|} \hline
Method					& Time ($s$)	\\ \hline 
MAP Viterbi				& 0.13			\\ \hline 
MAP Gradient			& 4.82			\\
MAP Hessian				& 0.51			\\ \hline
\end{tabular} 
\end{center} 
\caption[Frequency estimation]{Computation times comparison for frequency estimate.} 
\label{TableFreqEstTps} 
\end{table}

\begin{figure}[htbp]
\bcc
\btabu{cc}
\rotatebox{90}{\small \hspace{2.5cm} Frequency $\nu$} 
& \includegraphics[height=6cm]{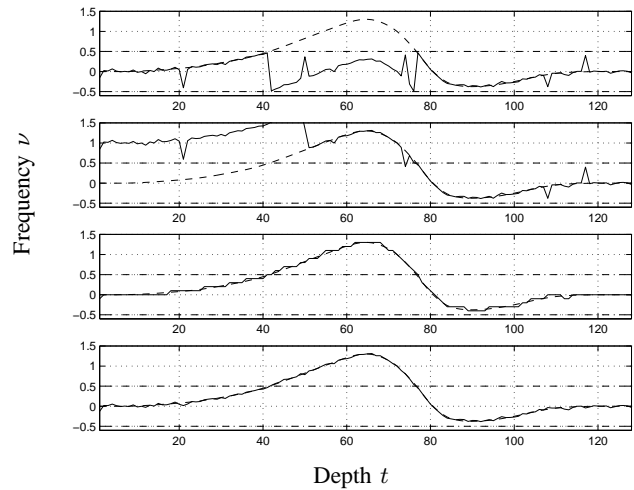} \\
& \small Depth $t$
\etabu
\ecc
\caption[Frequency estimation]{Comparison of frequency profile estimates. From top to
bottom: ML estimate (\ie periodogram maximizer), unwrapped ML estimate, Viterbi-MAP estimate
and Hessian-MAP estimate.}
\label{ResFreq}
\end{figure}

Fig.~\ref{ResFreq} illustrates typical results. The ML strategy:
\bit
\item[--] lacked robustness for two reasons: estimation was performed independently at each
depth and $N$ was small; 
\item[--] could not be corrected by an unwrap-like post-pro\-ces\-sing since the ML
solution was too rough (as already mentioned). 
\eit

For the regularized solution (also given in Fig.~\ref{ResFreq}), a simple qualitative
comparison with the reference led to three conclusions.
\bit
\item[--]   The estimated frequency sequence conformed much better to the true one. The
frequency sequence was more regular, since smoothness was introduced as a \prior feature.
\item[--] The estimated frequency sequence remained close to the true one even beyond the
usual Nyquist frequency. This was essentially due to the coherent accounting for the whole
set of data and smoothness of the frequency sequence. 
\item[--] The proposed strategy for estimating hyperparameters is adequate. A 
variation of 0.1 of the hyperparameters resulted in an almost imperceptible variation in the
estimated frequency sequence. This is especially important for qualifying the robustness of
the proposed method: the choice of $\rb$ offers relatively broad leeway and can be reliably
made. 

\eit

\section{Conclusion and perspectives} \label{Conclus}

This paper examines the problem of frequency tracking beyond the Nyquist frequency as it
occurs in Doppler imaging, when only short noisy data records are available. A solution is
proposed in the Bayesian framework based upon hidden Gauss-Markov models accounting for
\prior smoothness of the frequency sequence. We have developed a computationally efficient
combination of dynamic programming and a Hessian procedure  to calculate the maximum
\apost. The method is entirely unsupervised and uses an ML procedure based on an original
EM-based gradient procedure. The estimation of the ML hyperparameter is both formally
achievable and practically useful.

This new Bayesian method allows tracking beyond the usual Nyquist frequency, due to a
coherent statistical framework that includes the whole set of data plus smoothness \prior.
To our knowledge, this capability is an original contribution to the field of frequency
tracking.

Future work may include the extension to Gaussian DSP~\cite{Giovannelli94a}, to multiple
frequencies tracking~\cite{Streit90,Barret93}, and to the 2D problem. The latter and its
connection to 2D phase unwrapping~\cite{Ghiglia98,Servin98,Nico00} is presently being
investigated.

\appendix
\section{Amplitude marginalization}\label{AmpliMarg}

\subsection{Preliminary results}

This Section includes two useful results: for $\ub\in\eC^N$
\beqn
\Det{I_N + \ub \ub^\dag}			&=& 1 + \ub^\dag \ub \label{Determinant} \\
\pth{I_N + \ub^\dag \ub}\pmu		&=& I_N - \frac{\ub \ub^\dag}{1+\ub^\dag \ub}\label{InversionLemma}
\eeqn
where $I_N$ stands for the $N\times N$ identity matrix.

\subsection{Law for $(\yb_t | \nu_t)$}\label{ObservCondFreq}

Linearity of model~(\ref{ModelObserv}) \wrt amplitudes and assumptions for $a_t$ and
$\bb_t$, allow easy marginalization of $(\yb_t,a_t|\nu_t)$: $\yb_t|\nu_t$ is clearly a
zero-mean and Gaussian vector with covariance $R_t = r_a \zb(\nu_t) \zb(\nu_t)^\dag + r_b
I_N $. From relation~(\ref{Determinant}) and~(\ref{InversionLemma}) its determinant and
inverse reads:
\beqn 
R_t^{-1} &=& \frac{1}{r_b} I_N - \frac{\alpha}{N} \zb(\nu_t) \zb(\nu_t)^{\dag} \label{DetR} \\
\det R_t &=& r_b^{N-1} (r_b + Nr_a) \label{InvR}
\eeqn

\section{Proof of Proposition~\ref{PropDiffMAP}}\label{DemPropDiffMAP}

\subsection{Preliminary result}

The proposed proof is based on the decimal part function $D : \eR \longrightarrow [-0.5;
+0.5[$ defined by
\beq
\begin{cases}
D(x)=x ~\IF~ x\in[-0.5; +0.5), &\\
D \text{~is 1-periodic}, &\\
\end{cases}
\eeq
and the following straightforward properties
\beqn
&&D(x+k) = D(x), ~k\in\eZ\label{PropDec0}\\
&&|D(x)| \leqslant |x| \label{PropDec1}\\
&&|D(x)| \leqslant 1/2\label{PropDec2}\\
&&y=D(x) \Rightarrow \exists k\in\eZ \text{~such that~} y=x+k ~~\label{PropDec3}
\eeqn

\subsection{Proof of proposition}

Let us  define a frequency sequence $\nub$ (with $CLPL(\nub)<\infty$) which does not verify
Eq.~(\ref{EqPropDiffMAP}) of Proposition~\ref{PropDiffMAP}, \ie 
\beq \label{DemIneq}
\exists t_0\in\eN_{T-1}^* ~\WITH~ |\nu_{t_{0}+1} - \nu_{t_{0}}|>1/2 \,.
\eeq
Let us recursively build a new frequency sequence $\widetilde{\nub}$:
\beqn
\widetilde{\nu}_1		&=& \nu_1 \label{DemRec0}\\
\widetilde{\nu}_{t+1}	&=& \widetilde{\nu}_{t} + D(\nu_{t+1} - \widetilde{\nu}_t) ~\FOR~t=1,\dots,T-1
\label{DemRec}
\eeqn
and prove that Eq.~(\ref{EqPropDiffMAP}) of Proposition~\ref{PropDiffMAP} holds for
$\widetilde{\nub}$ and that the criterion $CLPL$ reduces from $\nub$ to $\widetilde{\nub}$:
\beqn
|\widetilde{\nu}_{t+1} - \widetilde{\nu}_t|	&\leqslant& 1/2 ~\FOR~t\in\eN_{T-1}^*  \,, \label{ADemDifNu}\\
CLPL(\widetilde{\nub})						&<& CLPL(\nub) \,.  \label{ADemDifCrit}
\eeqn

\indent$\bullet$~Relation~(\ref{ADemDifNu}) is straightforward: by Eq.~(\ref{DemRec}), one
can see
\beqx
\widetilde{\nu}_{t+1} - \widetilde{\nu}_t = D(\nu_{t+1} - \widetilde{\nu}_t) ~\FOR~t\in\eN_{T-1}^*
\eeqx
and hence, by Property~(\ref{PropDec2}):
\beqx
|\widetilde{\nu}_{t+1} - \widetilde{\nu}_t| \leqslant 1/2 ~\FOR~t\in\eN_{T-1}^* \,.
\eeqx

\indent$\bullet$~Proof of~(\ref{ADemDifCrit}) takes three steps, corresponding to each term
of CLPL~(\ref{CritPostLaw}). By Eq.~(\ref{DemRec0}-\ref{DemRec}) and
Property~(\ref{PropDec3}), one can see
\beq \label{NuEgalNuPlusK}
 \exists k_t\in\eZ \text{~~such that~~} \widetilde{\nu}_{t} = \nu_{t} + k_t ~\FOR~t\in\eN_{T}^* ~~\,,
\eeq
(with $k_1=0$) so, 
\beq \label{ProofEgalPerio}
P_t(\nu_t) = P_t(\widetilde{\nu}_t) ~\FOR~t\in\eN_{T}^* \,.
\eeq
By Eq.~(\ref{DemRec}) and~(\ref{NuEgalNuPlusK}) and invoking property~(\ref{PropDec0}) we
have 
\beqx
\widetilde{\nu}_{t+1} - \widetilde{\nu}_t = D(\nu_{t+1} - \widetilde{\nu}_t) = D(\nu_{t+1} - \nu_t)
\eeqx
hence, accounting for property~(\ref{PropDec1}):
\beq\label{ProofDiffNu1}
|\widetilde{\nu}_{t+1} - \widetilde{\nu}_t|  \leqslant |\nu_{t+1} - \nu_t| \,.
\eeq
Moreover, for $t=t_0$, we clearly have 
\beq\label{ProofDiffNu2}
|\widetilde{\nu}_{t_{0}+1} - \widetilde{\nu}_{t_{0}}|  < |\nu_{t_{0}+1} - \nu_{t_{0}}|
\eeq
thanks to hypothesis~(\ref{DemIneq}). Finally, we have:
\beq \label{ProofUnNu}
\unbb_K^\infty (\nu_1) = \unbb_K^\infty (\widetilde{\nu}_1)
\eeq
Collecting~(\ref{ProofEgalPerio}), (\ref{ProofDiffNu1}), (\ref{ProofDiffNu2})
and~(\ref{ProofUnNu}) proves~(\ref{ADemDifCrit}).

\section{HMC algorithms}

\subsection{The Viterbi algorithm}\label{AnnexViterbi}

\bcc
\begin{fminipage}[\LargeBox]
\bit
\item Pre-computations
\beqx
\barr{rcll}
\Dc(p,q)		&=& \lambda (\nu^p-\nu^q)^2	&~(p,q\in\eN^*_P)\\
\Lc(p,t)		&=& -P_t(\nu^p)					&~(p\in\eN^*_P,t\in\eN^*_T)
\earr
\eeqx
\item Initialization ($t=1$)
\beqx
\barr{rcll}
\Cc_1(p) &=& \Lc(p,1) \, \unbb_1^\infty(\nu^p) & (p\in\eN^*_P)
\earr
\eeqx
\item Iterations ($t=2,\dots,T$)
\beqx
\barr{rcll}
\widetilde{\Cc}_t(p,q)	&=& \Cc_{t-1}(q) + \Dc(p,q) + \Lc(p,t)		& (p,q\in\eN^*_P)\\
\Cc_t(p)					&=& \min_q \widetilde{\Cc}_t(p,q)					& (p\in\eN^*_P) \\
\Pc_t(p)					&=& \argmin_q \widetilde{\Cc}_t(p,q)				& (p\in\eN^*_P)
\earr
\eeqx
\item Termination ($t=T$)
\beqnx
\wh{p}_T	&=& \argmin_p \Cc_T(p)
\eeqnx
\item Back tracking ($t=T-1,\dots,1$)
\beqnx
\wh{p}_t	&=& \Pc_t(\wh{p}_{t+1})
\eeqnx
\eit
\end{fminipage}
\ecc

\subsection{The Forward algorithm}\label{AnnexForward}

\bcc
\begin{fminipage}[\LargeBox]
\bit
\item Initialization ($t=1$)
\beqx
\barr{rcll}
\overline{\Fc}_1(p)		&=& \Obb_1(p) \pbb(p)								& (p\in\eN^*_P) \\
\Nc_1							&=& \dispty\sum_{q=1}^P\overline{\Fc}_1(q)	&       \\
\Fc_1(p)                &=& \overline{\Fc}_1(p) / \Nc_1					& (p\in\eN^*_P)   
\earr
\eeqx
\item Iterations ($t=2,\dots,T$)
\beqx
\barr{rcll}
\overline{\Fc}_t(p)	&=& \Obb_t(p) \dispty\sum_{q=1}^P \Fc_{t-1}(p)  \Pbb(q,p)	& (p\in\eN^*_P)   \\
\Nc_t						&=& \dispty\sum_{q=1}^P \overline{\Fc}_t(q)						& \\
\Fc_t(p)					&=& \overline{\Fc}_t(p) / \Nc_t										& (p\in\eN^*_P)
\earr
\eeqx
\eit
\end{fminipage}
\ecc

\subsection{The Backward algorithm}\label{AnnexBackward}
\bcc
\begin{fminipage}[\LargeBox]
\bit
\item Initialization ($t=T$)
\beqx
\barr{rcll}
\overline{\Bc}_T(p)             &=& 1   & (p\in\eN^*_P)\\
\Bc_T(p)                        &=& 1   & (p\in\eN^*_P)
\earr
\eeqx
\item Iterations ($t=T-1,\dots,1$)
\beqx
\barr{rcll}
\overline{\Bc}_{t}(p)	&=& \dispty\sum_{q=1}^P \Obb_{t+1}(q) \Bc_{t+1}(p) \Pbb(p,q)	& (p\in\eN^*_P)   \\
\Bc_{t}(p)					&=& \overline{\Bc}_t(p) / \Nc_{t+1}										& (p\in\eN^*_P)
\earr
\eeqx
\eit
\end{fminipage}
\ecc

\section{Empirical estimation of hyperparameters }\label{HyperEmpiric}

This Section is devoted to the \textit{empirical} estimation of hyperparameters used as a
starting point in the maximization procedures of Section~\ref{ResultsHyper}. These
estimates are based on the correlation $r(n)$ of $\yb_t | \nu_t$ and easily shown to
verify: $r(0) = r_a+r_b \,, ~~\AND |r(1)| = r_a $, for all $t\in\eN_T^*$.  Empirical
estimates $\wh{r}(0)$ and $\wh{r}(1)$ are computed from the whole data set $\Yc$ and remain
robust, since $T$ is large (even if $N$ is small). Finally, one can compute $\wh{r}_a = 
|\wh{r}(1)|\,, \AND \wh{r}_b =  \wh{r}(0) - |\wh{r}(1)|$.

For $r_\nu$, the estimation is based on the ML estimate of the frequency sequence in each
range bin $t\in\eN_T^*$. The proposed empirical estimate of $r_\nu$ is naturally the
empirical variance of the differences between the ML frequencies. This procedure yields an
overestimated value for $r_\nu$. This result is expected, since the sequence of ML
frequencies varies greatly and has discontinuities, as mentioned above. Nevertheless, this
estimate is a suitable starting point for the maximization procedures of
Section~\ref{ResultsHyper}.



\begin{biography}
[{\includegraphics[width=1in,height=1.25in,clip,keepaspectratio]{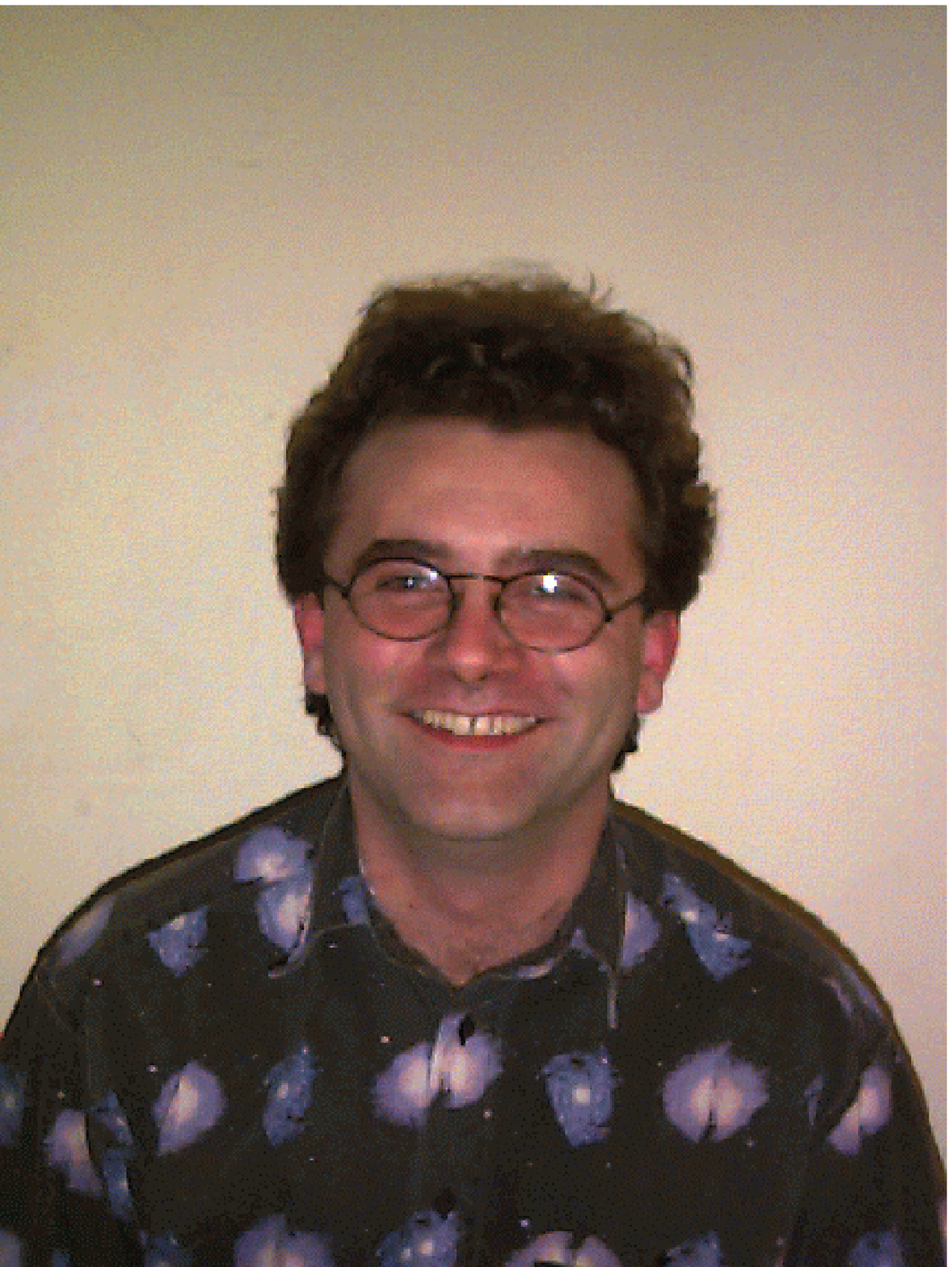}}]
{Jean-Fran\c{c}ois Giovannelli} was born in B\'e\-ziers, France, in 1966. He graduated from the \'Ecole Nationale Sup\'erieure de l'\'E\-l\-ec\-t\-ro\-ni\-que et de ses Applications in 1990. He received the Doctorat degree in physics at the Laboratoire des Signaux et Syst\`emes, Universit\'e de Paris-Sud, Orsay, France, in 1995. \\
\indent
He is presently assistant professor in the D\'epartement de Physique at Universit\'e de Paris-Sud. He is interested in regularization methods for inverse problems in signal and image processing, mainly in spectral  characterization. Application fields essentially concern radar and medical imaging.
\end{biography}

\begin{biography}
[{\includegraphics[width=1in,height=1.25in,clip,keepaspectratio]{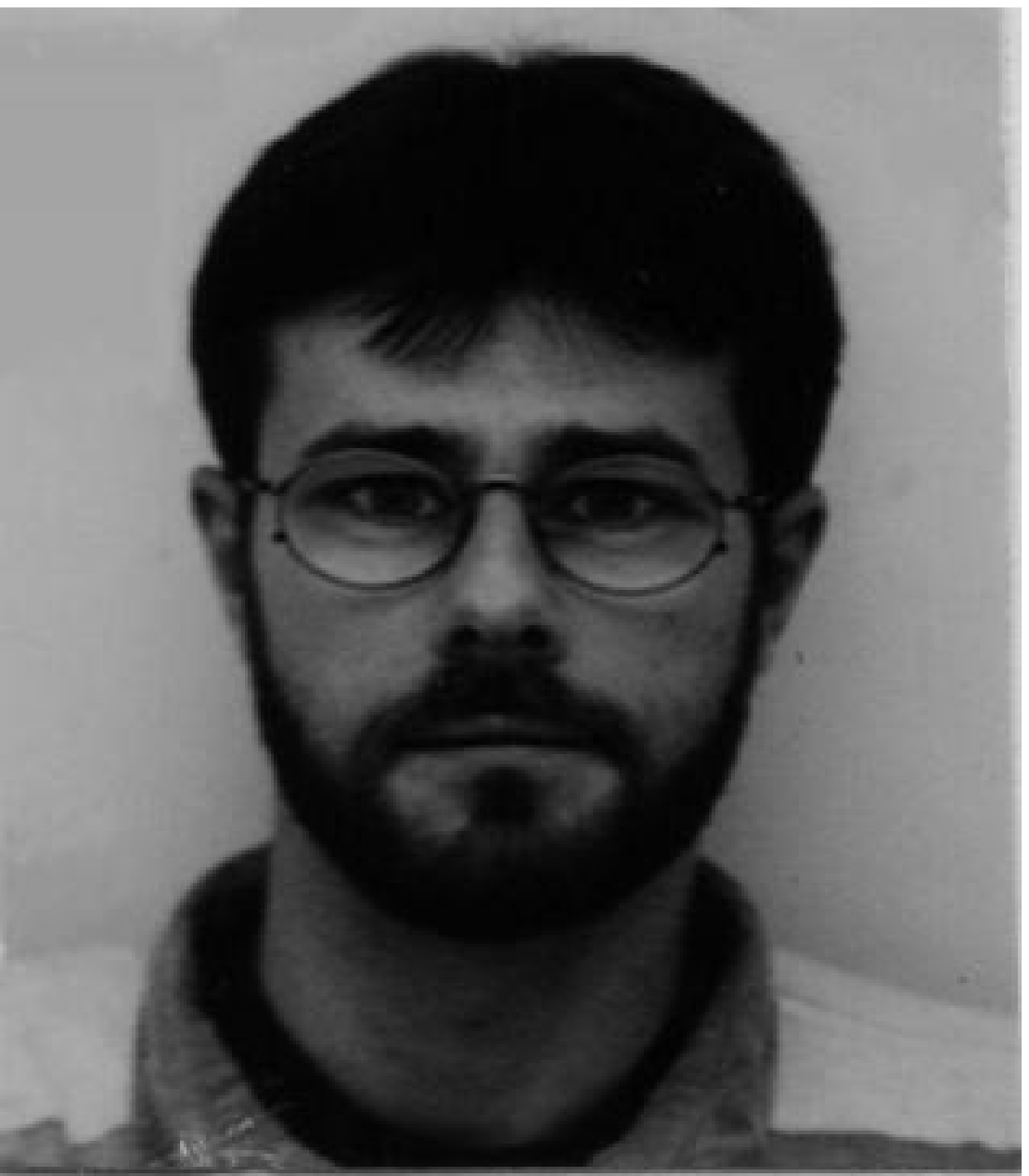}}]
{J\'er\^ome Idier} was born in France in 1966. He received the diploma degree in electrical engineering from the \'Ecole Sup\'erieure d'\'Electricit\'e in 1988 and the Ph.D. degree in physics from the Universit\'e de Paris-sud, Orsay, in 1991. Since 1991 he is with the Centre National de la Recherche Scientifique, assigned to the Laboratoire des Signaux et Syst\`emes. His major scientific interests are in probabilistic approaches to inverse problems for signal and image processing. 
\end{biography}

\begin{biography}
[{\includegraphics[width=1in,height=1.25in,clip,keepaspectratio]{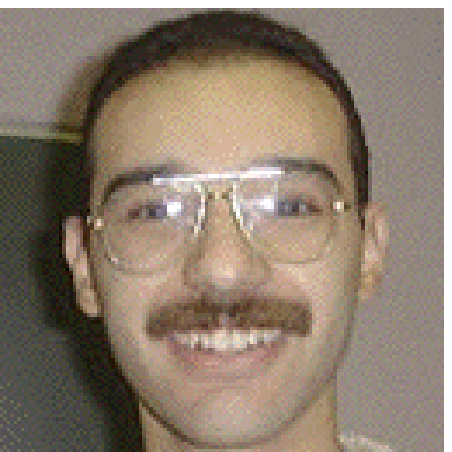}}]
{R\'edha Boubertakh} was born in Algiers, Algeria, in 1975. He received the diploma degree in electrical engineering from the \'Ecole Nationale Polytechnique d'Alger in 1996. He is currently pursuing the Ph.D. degree at the INSERM Unit 494, H\^opital Piti\'e-Salp\'etri\`ere, Paris, France. He is interested in signal and image processing mainly in the fields of magnetic resonance imaging.
\end{biography}

\begin{biography}
[{\includegraphics[width=1in,height=1.25in,clip,keepaspectratio]{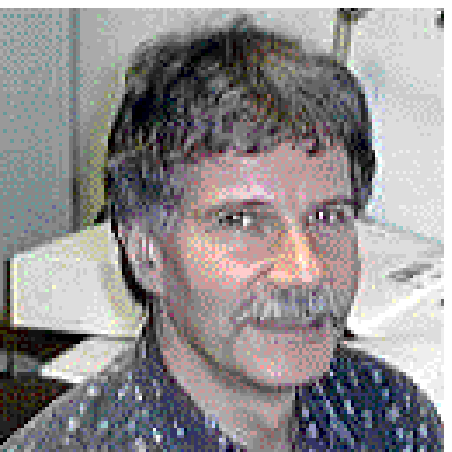}}]
{Alain Herment} was born in Paris, France, in 1948. He graduated from ISEP Engineering School of Paris in 1971. He received the Doctorat d'\'Etat degree in physics in 1984. 
\\ \indent
Initially, he worked as an engineer at the Centre National de la Recherche Scientifique. In 1977 he commenced a career as a researcher at the Institut National pour la Sant\'e et la Recherche M\'edicale (INSERM).  
\\ \indent
He is  currently in charge of the department of cardiovascular imaging at the INSERM Unit~66, H\^opital Piti\'e, Paris, France.  He is interested in signal and image processing for extracting morphological  and functional information from images sequences, mainly in the fields of  ultrasound investigations, X-ray CT, and digital angiography. 
\end{biography}

	\raggedbottom

\edoc